\documentclass[conference]{IEEEtran}

\usepackage{commath} 
\usepackage{cite}

\usepackage[pdftex]{graphicx}
\usepackage{array}

\usepackage{url} 
\usepackage{amsmath} 
\usepackage{nameref}

\usepackage{listings} 
\usepackage{booktabs}

\usepackage{alltt}
\usepackage[flushleft]{threeparttable}
\usepackage{multirow}
\usepackage{hhline}% http://ctan.org/pkg/hhline

\usepackage{enumitem}

\begin{document}
\title{A Machine Learning Driven IoT Solution \\ 
		for Noise Classification in Smart Cities}

\author{\IEEEauthorblockN{Yasser Alsouda}
	\IEEEauthorblockA{Dep. of Physics and Electrical Eng.\\
		 Linnaeus University\\
		 351 95 V\"{a}xj\"{o}, Sweden\\
		 Email: ya222ci@student.lnu.se}
	\and
	\IEEEauthorblockN{Sabri Pllana}
	\IEEEauthorblockA{Dep. of Computer Science and Media Tech.\\
		Linnaeus University\\
		351 95 V\"{a}xj\"{o}, Sweden\\
		Email: sabri.pllana@lnu.se}
    \and
    \IEEEauthorblockN{Arianit Kurti}
	\IEEEauthorblockA{RISE Interactive\\
		Research Institutes of Sweden\\
		602 33 Norrk\"{o}ping, Sweden\\
		Email: arianit.kurti@ri.se}
} 

%\IEEEspecialpapernotice{\footnotesize(ML-IoT 2018)}

\maketitle

%===========================================================================
\begin{abstract}

We present a machine learning based method for noise classification using a low-power and inexpensive IoT unit. We use Mel-frequency cepstral coefficients for audio feature extraction and supervised classification algorithms (that is, support vector machine and k-nearest neighbors) for noise classification. We evaluate our approach experimentally with a dataset of about 3000 sound samples grouped in eight sound classes (such as, car horn, jackhammer, or street music). We explore the parameter space of support vector machine and k-nearest neighbors algorithms to estimate the optimal parameter values for classification of sound samples in the dataset under study. We achieve a noise classification accuracy in the range 85\% -- 100\%. Training and testing of our k-nearest neighbors ($k = 1$) implementation on Raspberry Pi Zero W is less than a second for a dataset with features of more than 3000 sound samples.

\end{abstract}

\IEEEpeerreviewmaketitle

% % % % % % % % % % % % % % % % INTRODUCTION % % % % % % % % % % % % % % % %
\section{Introduction} 
\label{sec:intro}

About 85\% of Swedes live in urban areas and the quality of life and the health of citizens is affected by noise. Noise is any undesired environmental sound. The world health organization (WHO) recommends \cite{who} for good sleeping less than 30dB noise level in the bedroom and for teaching less than 35dB noise level in classroom. Recent studies \cite{poon} have found that exposure to noise pollution may increase the risk for health issues, such as, heart attack, obesity, impaired sleep, or depression. 

Following the Environmental Noise Directive (END) 2002/49/EC, each EU member state has to assess environmental noise and develop noise maps every five years. As sources of noise (such as, volume of traffic, construction sites, music and sport events) may change over time, there is a need for continuous monitoring of noise. Health damaging noise often occurs for only few minutes or hours, and it is not enough to measure the noise level every five years. Furthermore, \emph{the sound at the same dB level may be percepted as annoying noise or as a pleasant music}. Therefore, it is necessary to go beyond the state-of-the-art approaches that measure only the dB level \cite{rw:goetze,rw:tsao,rw:garcia} and in future we also \emph{identify the type of the noise}. For instance, it is important that the environment protection unit and law enforcement unit of a city know whether the noise is generated by a jackhammer at construction site or by a gun shot. The Internet of Things (IoT) is a promising technology for improving many domains, such as eHealth \cite{iot_ehealth,perez_fmec2018}, and it may be also used to address the issue of noise pollution in smart cities \cite{iot_sc}.

In this paper, we present an approach for noise classification in smart cities using machine learning on a low-power and inexpensive IoT unit. Mel-frequency cepstral coefficients (MFCC) are extracted as audio features and applied to two classifiers: support vector machine (SVM) and k-nearest neighbors (KNN). The evaluation of SVM and KNN with respect to accuracy and time is carried out on a Raspberry Pi Zero W. For evaluation we prepared a dataset of 3042 samples of environmental sounds  from UrbanSound8K \cite{UrbanSound} and Sound Events \cite{soundevent} in eight different classes (including gun shot, jackhammer, or street music). SVM classification performance is affected by parameters $\gamma$ and $C$, whereas parameter $k$ and minimum distance type (that is, Euclidean, Manhattan, or Chebyshev distance) influence the KNN performance. We explore the parameter space of SVM and KNN algorithms to estimate the optimal parameter values for classification of sound samples. The achieved noise classification accuracy is in the range 85\% -- 100\% and the time needed for training and testing of KNN model for $k = 1$ on Raspberry Pi Zero W is below one second. 

Major contributions of this paper include,
\begin{itemize}
	\item a machine learning approach for noise classification;
	\item implementation of our approach for noise classification on Raspberry Pi Zero W;
	\item experimental evaluation of our approach using a dataset of 3042 samples of environmental sounds;
    \item exploration of parameter space of KNN and SVM to estimate the best parameter values with respect to our sound samples dataset.
\end{itemize}

The rest of this paper is organized as follows. Section \ref{sec:background} gives an overview of machine learning and the Raspberry Pi platform. The proposed method for noise classification is described in Section \ref{sec:method}. Section \ref{sec:expeval} presents experimental evaluation of our approach, and Section \ref{sec:rewk} discusses the related work. The paper is concluded in Section \ref{sec:sum}.

% % % % % % % % % % % % % % % % BACKGROUND % % % % % % % % % % % % % % % %
\section{Background}
\label{sec:background}

% ------------------------------------------------------------------------
\subsection{Machine Learning}
\label{sec:ml}

Machine Learning is described by Mitchell \cite{mitchell97} as follows, "a computer program is said to learn from experience \emph{E} with respect to some class of tasks \emph{T} and performance measure \emph{P}, if its performance at tasks in \emph{T}, as measured by \emph{P}, improves with experience \emph{E}".

Commonly the supervised machine learning techniques are used for classification of data into different categories. Supervised learning means building a model based on known set of data (input and output) to predict the outputs of new data in the future. 
	In the midst of the diversity of classification algorithms, selecting the proper algorithm is not straightforward, since there is no perfect one that fits with all applications and there is always a trade-off between different model characteristics, such as: complexity, accuracy, memory usage, and speed of training.

% ------------------------------------------------------------------------
\subsection{Raspberry Pi and Mic-Hat}
\label{sec:hw}

\begin{figure}[ht]
	\centering
	\includegraphics[width=0.9\columnwidth]{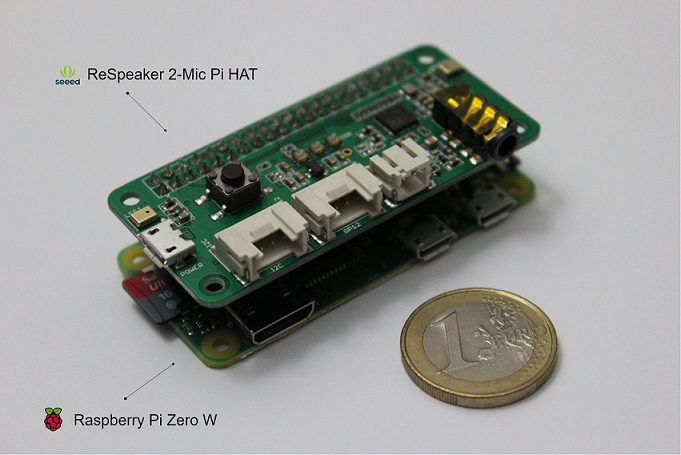}
	\caption{Noise classification hardware platform consists of a Raspberry Pi Zero W and a ReSpeaker 2-Mic Pi Hat.}
	\label{fig:hw}
\end{figure}

Figure \ref{fig:hw} depicts our hardware experimentation platform that comprises a Raspberry Pi Zero W and a ReSpeaker 2-Mic Pi HAT.

The Raspberry Pi \cite{rpi} is a low-power and low-cost single-board computer with a credit card size. It may be used as an affordable computer to learn programming or to build smart devices. A \emph{Raspberry Pi Zero W} with a Wi-Fi capability is used for our experiments. The Raspberry Pi Zero W (see Table \ref{tab:pi_properties}) comes with a single-core CPU running at 1GHz, 512MB of RAM, and costs only about \$10. 

We use for sound sensing a dual-mic array expansion board for Raspberry Pi called \emph{ReSpeaker 2-Mic Pi HAT} \cite{mic}. This board is developed based on WM8960 and has two microphones for collecting data and is designed to build flexible and powerful sound applications.

\begin{table}[ht]
\caption{Major properties of the Raspberry Pi Zero W}
\centering
\begin{tabular}{lc}
        \toprule
        Property		& Raspberry Pi Zero W \\
        \hline
        SOC				& Broadcom BCM2835 \\
		core			& 1 x ARM1176JZF-S, 1GHz \\   
        RAM		  		& 512MB  \\
        storage			& micro SD \\
        USB				& 1 x micro USB port \\        
        wireless LAN	& 802.11 b/g/n \\
        bluetooth		& 4.1 \\
        HDMI			& mini \\
        GPIO 			& 40 pins \\
        power (idle)	& 80mA (0.4W)\\
     	\bottomrule
\end{tabular}
\label{tab:pi_properties}
\end{table}

% % % % % % % % % % % % % % % % Method % % % % % % % % % % % % % % % %
\section{A Machine Learning Based Method for Noise Classification}
\label{sec:method}

\begin{figure}[ht]
	\centering
	\includegraphics[width=\columnwidth]{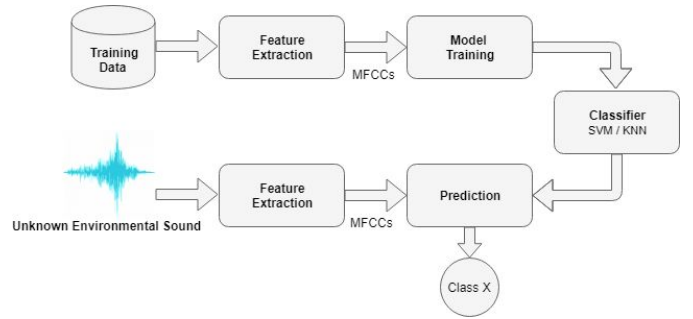}
	\caption{Our machine learning based approach for noise classification.}
	\label{fig:sys}
\end{figure}

In this section we describe our method for classification of noise using machine learning on Raspberry Pi. The proposed noise classification system is illustrated in Figure \ref{fig:sys}. MFCCs are extracted from a training dataset of sound samples to train SVM and KNN models that are used to predict the type of sensed environmental sounds.

% ------------------------------------------------------------------------
\subsection{Dataset}
\label{sec:dataset}

To investigate the performance of the system, we conduct experiments with eight different classes of environmental sounds: quietness, silence, car horn, children playing, gun shot, jackhammer, siren, and street music. For the purpose of this study we chose noise-relevant environmental sound clips from popular sound datasets, such as UrbanSound8K \cite{UrbanSound} and Sound Events \cite{soundevent}. The total dataset contains 3042 sound excerpts with length up to four seconds. Table \ref{tab:datasetinfo} provides the information about environmental sound samples that we use for experimentation.
    	
\begin{table}[ht]
\caption{Classes of sound samples in the dataset}
\centering
\begin{tabular}{lrc}
        \toprule
        Class   		&{Samples}	& {Duration}\\
        \hline
        Quietness		& 40   		& 02 min 00 sec\\
        %\hline
        Silence   	  	& 40   		& 02 min 00 sec\\
        %\hline
        Car horn		& 312  		& 14 min 38 sec\\
        %\hline
        Children playing& 560  		& 36 min 47 sec\\
        %\hline
        Gun shot  	  	& 235  		& 06 min 39 sec\\
        %\hline
        Jackhammer 	  	& 557  		& 32 min 34 sec\\
        %\hline
        Siren  	  		& 662  		& 43 min 17 sec\\
        %\hline
        Street music  	& 636  		& 42 min 24 sec\\
     	\hline
        Total			& 3042		& 2 hrs 0 min 19 sec\\
        \bottomrule
\end{tabular}
\label{tab:datasetinfo}
\end{table}

% ------------------------------------------------------------------------
\subsection{Feature Extraction}
\label{features}

\begin{figure}[ht]
			\centering
			\includegraphics[width=\columnwidth]{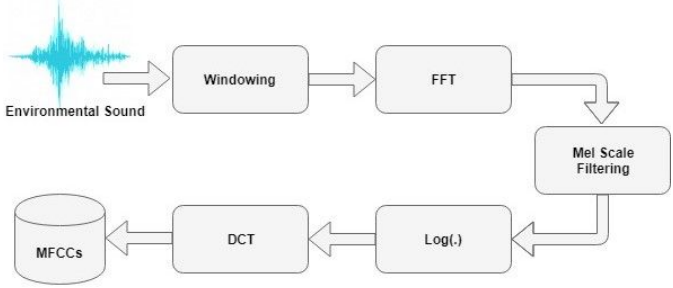}
			\caption{The procedure for generating MFCCs of environmental sounds.}
			\label{fig:mfcc}
		\end{figure}

Features extraction is the first step in an automatic sound classification system. MFCCs~\cite{sahidullah} are a well-known feature set and are widely used in the area of sound classification because they are well-correlated to what the human can hear. MFCCs are obtained using the procedure depicted in Figure \ref{fig:mfcc}.
	
Foote \cite{foote} proposes the use of the first 12 MFCCs plus an energy term as sound features. In this paper, we computed the first 12 MFCCs of all frames of the entire signal and appended the frame energy to each feature vector, thus each audio signal is transformed into a sequence of 13-dimensional feature vector.
	
% ------------------------------------------------------------------------	
\subsection{Classification}
\label{sec:classification}

In this section we examine two supervised classification methods: support vector machine  and k-nearest neighbors. 

% ------------------------------------------------------------------------	
\subsubsection{Support Vector Machines (SVM)}
\label{sec:svm}

SVM \cite{Bishop2006} is a popular supervised algorithm mostly used for solving classification problems. The main goal of the SVM algorithm is to design a model that finds the optimal hyperplane that can separate all training data into two classes. There may be many hyperplanes that separate all the training data correctly, but the best choice will be the hyperplane that leaves the maximum margin, which is defined as the distance between the hyperplane and the closest samples. Those closest samples are called the support vectors.

\begin{figure}[ht]
			\centering
			\includegraphics[width=0.7\columnwidth]{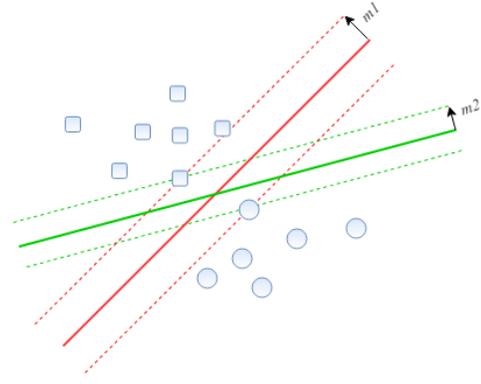}
			\caption{An illustration of SVM for a 2-class classification problem.}
			\label{fig:svm_2class}
		\end{figure}

Considering the example of two linearly separable classes (circles and squares) shown in Figure \ref{fig:svm_2class}, both hyperplanes ($one$ and $two$) can classify all the training instances correctly, but the best hyperplane is $one$ since it has a greater margin ($m_1 > m_2$).

	When the data is nonlinearly separable, the nonlinear classifier can by created by applying the kernel trick \cite{Theo2009}. Using the kernel trick, the non-separable problem can be converted to a separable problem using kernel functions that transform low dimensional input space to high dimensional space. Selecting the appropriate kernel and its parameters has a significant impact on the SVM classifier. Another important parameter for the SVM classifier is the soft margin parameter $C$, which controls the trade-off between the simplicity of the decision boundary and the misclassification penalty of the training points. A low value of $C$ makes the classifier tolerant with misclassified data points (that is, \emph{smooth decision boundary}), while a high value of $C$ makes it aiming to a perfect classification of the training points (that is, \emph{complex boundary decision}).
   
    One of the kernel functions that is commonly used in SVM classification is the radial basis function (RBF). The RBF kernel on two feature vectors (\textit{x} and \textit{x'}) is expressed by Equation \ref{eq:rbf}.
   
        \begin{equation}
        \label{eq:rbf}
        	K(x, x') = \exp{(-\frac{\|x - x'\|^{2}}{2\sigma^2})} = \exp{(-\gamma \|x - x'\|^{2})}
        \end{equation}
        
	The RBF parameter $\gamma$ determines the influence of the training data points on determining the exact shape of the decision boundary. With a high value of $\gamma$ the details of the decision boundary are determined only by the closest points, while for a low value of $\gamma$ even the faraway points are considered in drawing the decision boundary.
    
In this paper, we explore the effect of parameters $\gamma$ and $C$ on SVM model with respect to our dataset of sound samples. 

% ------------------------------------------------------------------------	
\subsubsection{K-Nearest Neighbors (KNN)}
\label{sec:knn}
    
KNN is one of the simplest machine learning algorithms used for classification. The KNN works based on the minimum distance (such as, Euclidean distance) between the test point and all training points. The class of the test point is then determined by the most frequent class of the $k$ nearest neighbors to the test point. Commonly used distances include,
    \newline	        
	
		\begin{itemize}
			\item \emph{Euclidean distance}: $d$(q, p) = $\sqrt{\sum_{i=1}^{n} (q_i - p_i)^{2}}$
            \newline
			\item \emph{Manhattan distance}: $d$(q, p) = $\sum_{i=1}^{n} |q_i - p_i|$
            \newline
			\item \emph{Chebyshev distance}: $d$(q, p) = $\max_{i} (|q_i - p_i|)$ 
            \newline
		\end{itemize}
            
\begin{figure}[ht]
			\centering
			\includegraphics[width=0.5\columnwidth]{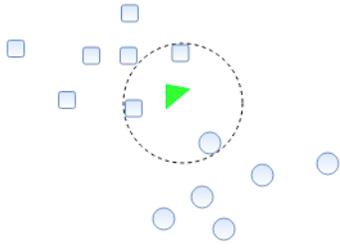}
			\caption{An illustration of KNN for a 2-class classification problem for $k = 3$.}
			\label{fig:knn_2class}
		\end{figure}

The KNN classifier is illustrated with an example in Figure \ref{fig:knn_2class}. Two classes are represented with \emph{squares} and \emph{circles} and the aim of the KNN algorithm is to predict the correct class of the \emph{triangle}. Suppose $k = 3$, then the model will find three nearest neighbors of triangle. To predict the correct class of the triangle, the algorithm can achieve its aim by finding three nearest neighbors of the triangle and the most frequent element determines the class of the \emph{triangle}, which is the class of \emph{squares} in this case. 

The KNN algorithm needs a significant amount of memory to run, since it requires all the training data to make a prediction.

% % % % % % % % % % % % % EXPERIMENTAL EVALUATION % % % % % % % % % % % % %
\section{Experimental Evaluation}
\label{sec:expeval}
	
In this section, we investigate the performance of SVM and KNN on eight different classes of environmental sounds: quietness, silence, car horn, children playing, gun shot, jackhammer, siren, street music. For training the models we use a dataset of 3042 samples of environmental sounds (see Table \ref{tab:datasetinfo}). We divide the dataset arbitrary into two sub-sets: 75\% are used for training and 25\% for testing. All experiments are repeated 20 times with different sub-sets and the obtained results are averaged. We have implemented all algorithms in Python using open source packages for machine learning and audio analysis (that is, scikit-learn \cite{scikit} and librosa \cite{librosa}).
    
% ------------------------------------------------------------------------	
\subsection{SVM Parameter Space Exploration}
\label{exp_svm}
    
\begin{figure}
    \centering
     \includegraphics[width=0.9\columnwidth]{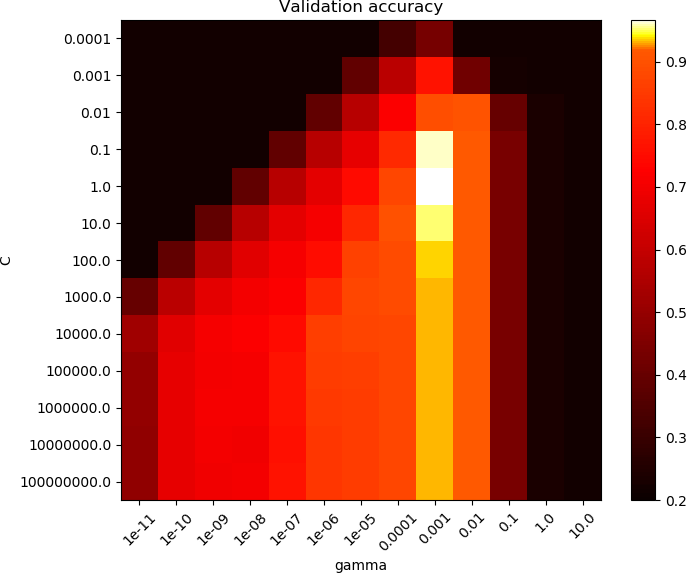}
     \caption{Heat map of the SVM validation accuracy as a function of $\gamma$ and $C$.}
     \label{fig:svm_valacc}
\end{figure}

\begin{figure}[ht]
			\centering
			\includegraphics[width=0.9\columnwidth]{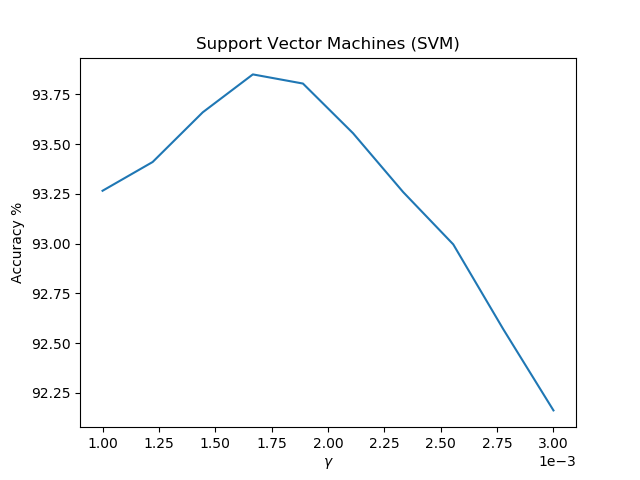}
			\caption{The effect of the parameter $\gamma$ on the performance of the SVM classifier.}
			\label{fig:svm_gm}
		\end{figure}

\begin{figure}[ht]
			\centering
			\includegraphics[width=0.9\columnwidth]{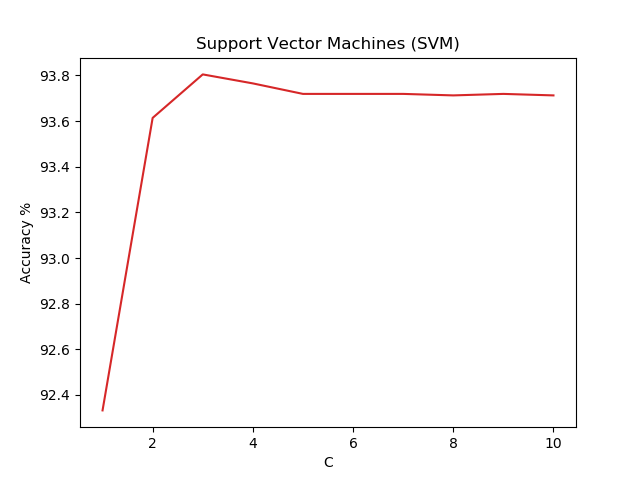}
			\caption{The effect of the parameter $C$ on the performance of the SVM classifier.}
			\label{fig:svm_C}
		\end{figure}

    To optimize the performance of SVM, the grid search is used to select the best combination of the parameters $\gamma$ and $C$ for the RBF kernel. To explore the SVM's cross-validation accuracy, we plot the heat map depicted in Figure \ref{fig:svm_valacc} as a function of $\gamma$ and $C$, where $\gamma$ $\epsilon$ $\{ 10^{-11} - 10^1 \}$ and $C$ $\epsilon$ $\{10^{-4} - 10^8\}$. Table \ref{tab:exp_svm} shows the SVM model accuracy [\%] for various values of $\gamma$ and $C$ parameters. After evaluating the model, we achieved a 93.87\% accuracy for $\gamma = 0.00167$ and $C = 3$, as shown in Figure \ref{fig:svm_gm} and Figure \ref{fig:svm_C}.
    
        \begin{table}[ht]
        \caption{Accuracy [\%] of SVM}
        \centering
    	\begin{tabular}{l|ccccc}
        \toprule
         & \multicolumn{4}{c}{\textsf{$\gamma$}}\\
        \cline{2-5}
         \textit{C}& 0.0001   	& 0.00167		& 0.01		& 0.1 \\
       	\hline
        0.1 & 64.14				& 67.07			& 22.96		& 21.18	\\
        
         1	& 79.19   			& 92.31 		& 72.66		& 29.88	\\
        
         3  & 82.85				& 93.87			& 75.22		& 31.53	\\
        
         5	& 84.40  			& 93.86			& 75.21		& 31.53	\\
        
         10	& 85.90  			& 93.83			& 75.19		& 31.53	\\
         
        100 & 89.24  			& 93.70			& 75.18		& 31.54	\\
        
     	\bottomrule
    	\end{tabular}
        \label{tab:exp_svm}
        \end{table}
        
% ------------------------------------------------------------------------	
	\subsection{KNN Parameter Space Exploration}
    \label{exp_knn}
    
\begin{figure}[ht]
			\centering
			\includegraphics[width=0.9\columnwidth]{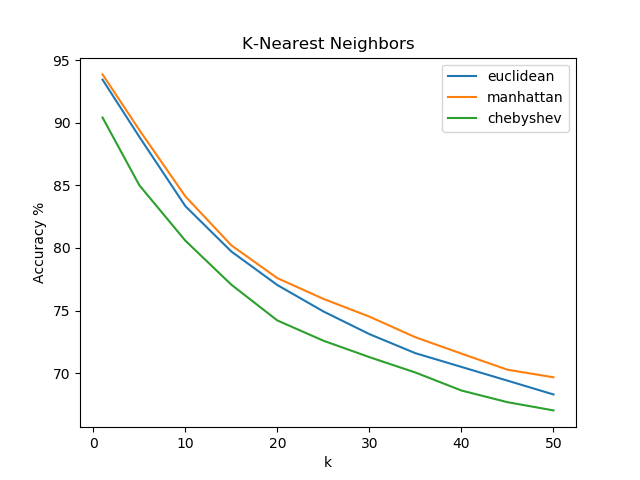}
			\caption{Performance of the KNN classifier for various values of nearest neighbors $k$ and Euclidean, Manhattan, and Chebyshev distances.}
			\label{fig:knn_k_dis}
		\end{figure}

For KNN classifier we examine the influence of parameter $k$, the Euclidean distance, Manhattan distance, and the Chebyshev distance (Section \ref{sec:knn}). Figure \ref{fig:knn_k_dis} illustrates the classification accuracy of KNN for various values of $k$ for each kind of distance. Table \ref{tab:exp_knn} presents the results for the KNN accuracy, where the $Manhattan$ distance and $k = 1$ proved to be the best parameters with sound type recognition accuracy of 93.88$\%$.
    
        \begin{table}[ht]
        \caption{Accuracy [\%] of KNN}
        \centering
    	\begin{tabular}{l|ccc}
        \toprule
        	& \multicolumn{3}{c}{\textit{Distance}}\\
            \cline{2-4}
        		\textit{k}		& Euclidean   		& Manhattan		& Chebyshev\\
        \hline
         1	& 93.46   			& 93.88 		& 90.43\\
        
         5  & 88.88				& 89.42			& 85.01\\
        
         10	& 83.34  			& 84.13			& 80.58\\
         
         50 & 68.20				& 69.66			& 67.01\\
        
     	\toprule
    	\end{tabular}
        \label{tab:exp_knn}
        \end{table}

% ------------------------------------------------------------------------	
	\subsection{Performance of SVM and KNN}
    \label{exp_comp}
    
\begin{figure}[ht]
			\centering
			\includegraphics[width=\columnwidth]{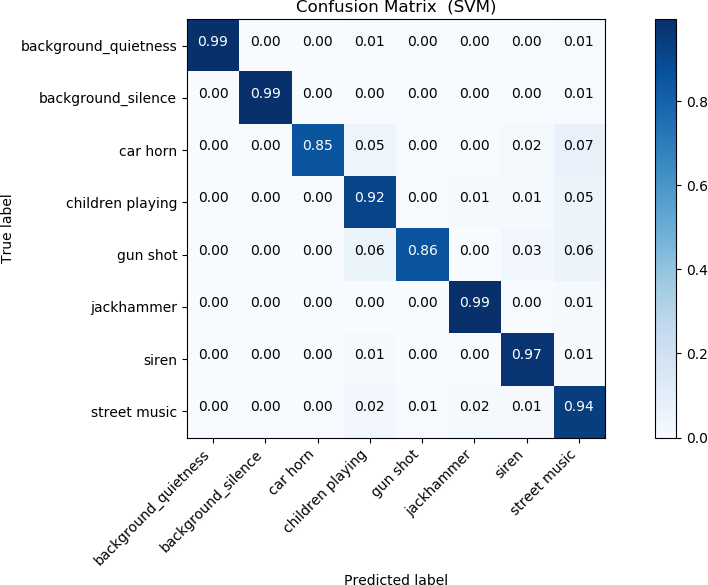}
			\caption{SVM-based classification of noise.}
			\label{fig:cm_svm}
		\end{figure}

\begin{figure}[ht]
			\centering
			\includegraphics[width=\columnwidth]{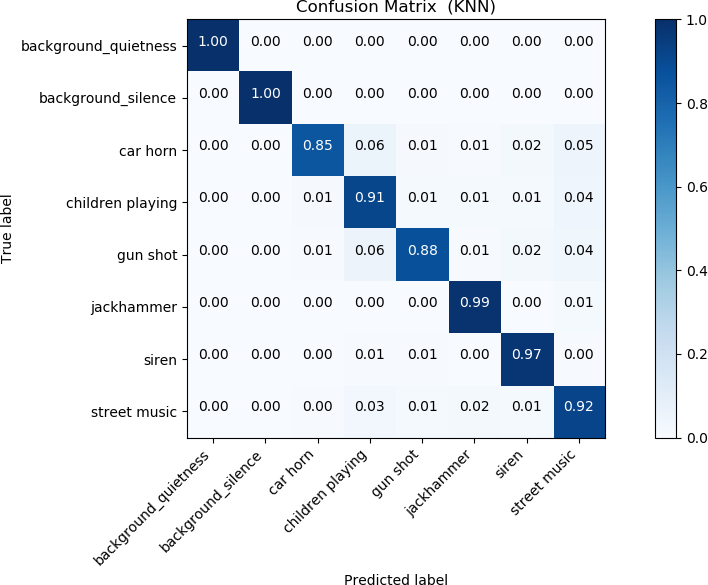}
			\caption{KNN-based classification of noise.}
			\label{fig:cm_knn}
		\end{figure}

In this section we present the performance of SVM and KNN with respect to classification accuracy and time that is needed for training and testing. To examine the accuracy of each model we plot the confusion matrix that compares the predicted classes with the true noise classes. Figure \ref{fig:cm_svm} and Figure \ref{fig:cm_knn} illustrate the confusion matrices of SVM and KNN, respectively, while Table \ref{tab:time_svm} and Table \ref{tab:time_knn} present the time performance of SVM and KNN, respectively, during training and testing on the Raspberry Pi Zero W.
    
\begin{table}
    	\caption{Time [seconds] for training and testing of SVM model on Pi Zero W. The time for feature extraction is not included.}
        \centering
        \begin{tabular}{c|cccccccc}
        \toprule
        & \multicolumn{8}{c}{$\gamma$}\\
        \cline{2-9}
        \textit{C} & \multicolumn{2}{c}{0.0001} & \multicolumn{2}{c}{0.00167} & \multicolumn{2}{c}{0.01} & \multicolumn{2}{c}{0.1} \\
         	& Train & Test & Train & Test & Train & Test & Train & Test \\ \midrule
        0.1 & 8.03  & 2.37 & 11.90 & 2.59 & 21.98 & 2.87 & 31.56 & 4.64 \\
          1 & 5.00  & 1.90 & 11.93 & 1.98 & 26.37 & 2.58 & 33.00 & 4.56 \\
          3 & 4.36  & 1.63 & 12.29 & 1.99 & 26.70 & 2.65 & 33.42 & 4.50 \\
          5 & 4.50  & 1.62 & 12.44 & 1.99 & 26.76 & 2.56 & 33.36 & 4.51 \\
         10 & 4.29  & 1.41 & 12.33 & 1.98 & 26.85 & 2.56 & 35.32 & 4.77 \\
        100 & 5.50  & 1.17 & 12.29 & 1.98 & 26.59 & 2.58 & 34.24 & 4.65 \\ 
         \bottomrule
\end{tabular}
        \label{tab:time_svm}
    	\end{table}
        
\begin{table}
    	\caption{Time [seconds] for training and testing of KNN model on Pi Zero W. The time for feature extraction is not included.}
        \centering
        \begin{tabular}{c|cccccc}
        \toprule
        & \multicolumn{6}{c}{\textit{Distance}}\\
        \cline{2-7}
        \textit{k} & \multicolumn{2}{c}{\textit{Euclidean}} & \multicolumn{2}{c}{\textit{Manhattan}} & \multicolumn{2}{c}{\textit{Chebyshev}} \\
         &Train &  Test & Train & Test &  Train &  Test \\ \midrule
 	   1 & 0.05	& 0.21	& 0.05	& 0.5  &  0.05  &  0.14 \\
       5 & 0.05 & 0.37  & 0.05  & 0.92 &  0.05  &  0.24 \\
      10 & 0.05 & 0.47  & 0.05  & 1.15 &  0.05  &  0.31 \\
     100 & 0.05 & 0.80  & 0.05  & 1.71 &  0.05  &  0.57 \\
         \bottomrule
\end{tabular}
        
        \label{tab:time_knn}
    	\end{table}

% % % % % % % % % % % % % % % % Related Work % % % % % % % % % % % % % % % %
\section{Related Work}
\label{sec:rewk}

In this section we discuss the related work with respect to IoT solutions for noise measurement and machine learning methods for sound classification.

% ------------------------------------------------------------------------	
	\subsection{IoT Solutions for Noise Measurement}
    
    Goetze et al \cite{rw:goetze} provide an overview of a platform for distributed urban noise measurement, which is part of an ongoing German research project called \emph{StadtLärm}. A wireless distributed network of audio sensors based on quad-core ARM BCM2837 SoC was employed to receive urban noise signals, pre-process the obtained audio data and send it to a central unit for data storage and performing higher-level audio processing. A final stage of web application was used for visualization and administration of both processed and unprocessed audio data. The authors in \cite{rw:tsao} used Ameba RTL 8195AM and Ameba 8170AF as IoT platforms to implement a distributed sensing system for visualization of the noise pollution. In \cite{rw:garcia}, two hardware alternatives, Raspberry Pi platform and Tmote-Invent nodes, were evaluated in terms of their cost and feasibility for analyzing urban noise and measuring the psycho-acoustic metrics according to the Zwicker's annoyance model. 
    
    In contrast to related work, our approach is not concerned with measuring the noise level in dB using IoT, but with determining the type of noise (for instance, a jackhammer or gun shot). 

% ------------------------------------------------------------------------	
	\subsection{Machine Learning Methods for Sound Classification}
    
     In \cite{rw:wang}, a combination of two supervised classification methods, SVM and KNN, were used as a hybrid classifier with MPEG-7 audio low-level descriptor as the sound feature. The experiments were conducted on 12 classes of sounds. Khunarasal et al \cite{rw:khunarsal} proposed an approach to classify 20 different classes of very short time sounds. The study investigated various audio features (e.g., MFCC, MP, LPC and Spectrogram) along with KNN and neural network. 
     
     We complement the related work, with a study of noise classification on a low-power and inexpensive device, that is the Raspberry Pi Zero W. 

% % % % % % % % % % % % % % % % Conclusion % % % % % % % % % % % % % % % %
\section{Summary}
\label{sec:sum}

 We have presented a machine learning approach for noise classification. Our method uses MFCC for audio feature extraction and supervised classification algorithms (that is, SVM or KNN) for noise classification. We implemented our approach using Raspberry Pi Zero W that is a low-power and inexpensive hardware unit. We observed in our experiments with various environment sounds (such as, car horn, jackhammer, or street music) that KNN and SVM provide high noise classification accuracy that is in the range 85\% -- 100\%. Experiments with various values of parameter $k$, which determines the number of nearest data neighbors, indicate that the accuracy of KNN decreases with the increase of $k$. Experiments with various values of parameter $C$, which determines misclassification penalty, indicate that SVM had the highest accuracy for $C = 3$ for our dataset. The dataset used in our experiments contains features of about 3000 sound samples and training and testing of KNN ($k = 1$) on Pi Zero W took a fraction of second.  
    
Future work will investigate usefulness of our solution for a large number of Raspberry Pi devices in an environment that combines features of the Edge and Cloud computing systems. 

% % % % % % % % % % % % % % % % References % % % % % % % % % % % % % % % %

\end{document}